\documentclass[letterpaper,journal]{IEEEtran}
\usepackage{amsmath}   
\usepackage{amssymb}   
\usepackage{braket}    
\usepackage{graphicx}
\usepackage{algorithm}      
\usepackage{algorithmicx}   
\usepackage{algpseudocode}  

\begin{document}

\pagestyle{empty}
\thispagestyle{empty}


\title{Proof-of-Spiking-Neurons(PoSN): Neuromorphic Consensus for Next-Generation Blockchains}

\author{M.Z. Haider$^{1}$, M.U. Ghouri$^{2}$, Tayyaba Noreen$^{1}$, M. Salman$^{3}$ \\
$^{1}$Department of Software Engineering(ÉTS), Université du Québec, Canada \\
$^{2}$Department Of Computational Sciences,The University of Faisalabad, Pakistan
\\
$^{3}$Department of Computer Science, SZABIST University
}

\maketitle
\thispagestyle{empty}

\begin{abstract}
Blockchain systems face persistent challenges of scalability, latency, and energy inefficiency. Existing consensus protocols such as Proof-of-Work (PoW) and Proof-of-Stake (PoS) either consume excessive resources or risk centralization. This paper proposes \textit{Proof-of-Spiking-Neurons (PoSN)}, a neuromorphic consensus protocol inspired by spiking neural networks. PoSN encodes transactions as spike trains, elects leaders through competitive firing dynamics, and finalizes blocks via neural synchronization, enabling parallel and event-driven consensus with minimal energy overhead. A hybrid system architecture is implemented on neuromorphic platforms, supported by simulation frameworks such as Nengo and PyNN. Experimental results show significant gains in energy efficiency, throughput, and convergence compared to PoB and PoR. PoSN establishes a foundation for sustainable, adaptive blockchains suitable for IoT, edge, and large-scale distributed systems.
\end{abstract}


\section{Introduction}

Blockchain has become a core technology for secure and transparent systems in finance, supply chain, and healthcare. Despite its adoption, scalability remains a challenge: public blockchains like Bitcoin and Ethereum process only a few TPS compared to thousands in centralized systems such as Visa~\cite{ali2022sustainable,cao2022scalability}. This gap underscores the need for more efficient consensus mechanisms without sacrificing decentralization or security \cite{xu2021scaling}. Traditional consensus mechanisms such as Proof-of-Work (PoW) ensure strong security but consume excessive energy, with Bitcoin alone using as much as a medium-sized country~\cite{zhang2022energy}. Proof-of-Stake (PoS) reduces waste but raises concerns over centralization and stake accumulation attacks~\cite{saleh2021blockchain,chauhan2023energy}. BFT-based protocols improve efficiency but scale poorly in large networks due to communication overhead~\cite{cachin2021blockchain}. These limitations motivate exploring alternative paradigms that deliver high throughput, low energy cost, and strong security \cite{haider2025towards}.

Recent work explores energy-efficient consensus for blockchain\cite{gharavi2024post}. Lightweight protocols for IoT and edge aim to balance energy use and fault tolerance~\cite{zhang2022energy,liu2023green}, while AI-enhanced frameworks support dynamic leader election, load balancing, and anomaly detection~\cite{qiu2023ai,huang2023blockchain}. Yet, these approaches remain limited by von Neumann architectures, which execute sequentially and struggle with the parallel, real-time demands of large-scale blockchains\cite{varadarajan2024innovative}. Neuromorphic computing offers a brain-inspired paradigm that enables energy efficiency, parallelism, and adaptivity\cite{paykari2025enhancing}. Processors such as Intel Loihi, SpiNNaker, and IBM TrueNorth use spiking neural networks (SNNs) for event-driven computation. Unlike CPUs and GPUs, they perform massively parallel operations with far lower power consumption, making them promising for edge and distributed systems~\cite{schuman2022opportunities,roy2023neuromorphic}. Their efficiency and adaptability align well with blockchain consensus needs, which require both security and responsiveness under resource constraints\cite{motaqy2024chainboost}.

\begin{figure}[htbp]
    \centering
    \includegraphics[width=0.9\linewidth]{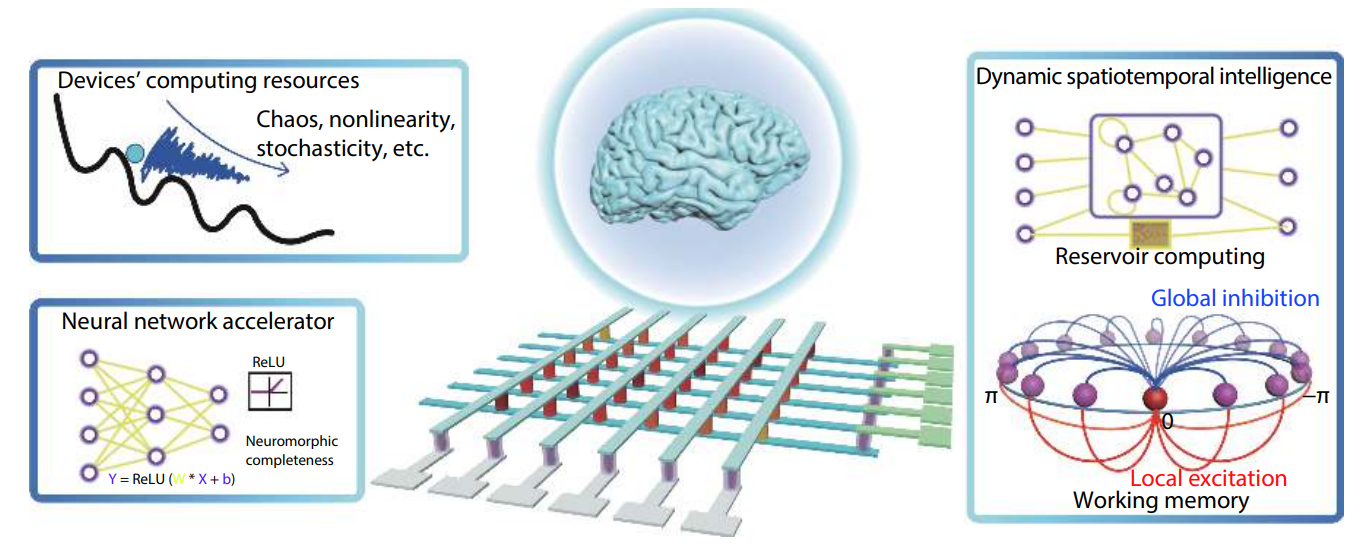}
    \caption{Roadmap of neuromorphic computing \cite{wang2021embracing}.}
    \label{fig:throughput-comparison}
\end{figure}

We propose \textit{Proof-of-Spiking-Neurons (PoSN)}, a neuromorphic consensus mechanism that leverages spiking neural dynamics for secure, fair, and energy-efficient block validation. Transactions are encoded as spike trains, leader election emerges from spiking activity, and block finalization from synchronized neural firing. Unlike PoW’s brute-force computations or PoS’s risk of stake centralization, PoSN introduces a biologically inspired, low-power consensus. To our knowledge, this is the first consensus protocol designed and prototyped for neuromorphic hardware in blockchain systems. By bridging neuromorphic computing with blockchain, this work opens a new research direction in consensus design. PoSN represents a step toward sustainable and intelligent blockchains that can scale in environments where traditional approaches fail, setting the stage for future bio-inspired distributed ledger systems. The key contributions of this work are:  
\begin{enumerate}
    \item Proposal of \textit{PoSN}, a spiking neural network-based consensus replacing cryptographic puzzles and stake in leader election.
    \item Prototype implementation on neuromorphic simulators (Nengo, PyNN) benchmarked for latency, and throughput against PoB and PoR.
    \item Security and scalability analysis demonstrating PoSN’s resilience to Sybil and Byzantine attacks and suitability for IoT and edge systems.
\end{enumerate}

The remaining sections are organized as follows. Section~\ref{sec: background} presents the background and related work. Section~\ref{sec: protocol} introduces our proposed protocol,Section~\ref{sec: evaluation} reports the evaluation results, and Section~\ref{sec: conclusion} concludes the paper.

\section{Related Work}
\label{sec: background}
Consensus underpins blockchain systems, enabling nodes to agree on transaction validity and ledger state \cite{upadhyay2024need}. Classical mechanisms like Proof-of-Work (PoW), Proof-of-Stake (PoS), and Byzantine Fault Tolerance (BFT) are well-studied but face scalability and efficiency limitations in large networks \cite{jia2025elm}. PoW, while highly secure and decentralized, has been heavily criticized for its excessive energy consumption and low throughput, consuming power comparable to medium-sized countries~\cite{zhang2022energy}. PoS emerged as a greener alternative, reducing energy consumption by replacing computational puzzles with stake-based selection, yet it introduces centralization risks and wealth concentration concerns~\cite{saleh2021blockchain}. Similarly, BFT-based protocols achieve lower latency and deterministic finality but scale poorly due to their quadratic communication complexity, restricting their use in large, decentralized environments~\cite{cachin2021blockchain}. In recent years, 
Proof-of-Brain (PoB) \cite{rodinko2024decentralized} and Proof-of-Randomness (PoR)\cite{raikwar2022sok} have emerged as alternative consensus paradigms addressing fairness and efficiency in decentralized systems. PoB introduces a human-centric validation model based on cognitive or contribution-based scoring to promote equitable participation, whereas PoR leverages cryptographic randomness to ensure unbiased leader selection and reduce manipulation in block generation. Green consensus approaches attempt to minimize redundant computations and reduce hardware demands, particularly in Internet of Things (IoT) and edge computing environments~\cite{liu2023green}. Hybrid protocols that combine PoS with BFT or committee-based leader selection have been shown to balance efficiency with fault tolerance, but challenges in fairness and resistance to adversarial manipulation remain unresolved~\cite{cao2022scalability,chauhan2023energy}. Despite these advancements, there is still no consensus mechanism that achieves the trifecta of scalability, strong decentralization, and low energy consumption simultaneously~\cite{haider2025v}.

Parallel to developments in blockchain, neuromorphic computing has emerged as a novel paradigm that mimics biological brains through Spiking Neural Networks (SNNs)\cite{nunes2022spiking}. Neuromorphic processors such as Intel Loihi, SpiNNaker\cite{gonzalez2024spinnaker2}, and IBM TrueNorth execute computations in an event-driven and massively parallel manner, enabling orders-of-magnitude energy savings compared to von Neumann architectures~\cite{schuman2022opportunities,roy2023neuromorphic}. These chips have demonstrated success in real-time decision-making, temporal sequence recognition, and adaptive control, which are computational characteristics that align closely with blockchain consensus requirements\cite{yang2025trust}. In particular, their ability to support distributed and low-power operations makes them especially suitable for blockchain applications in IoT and edge systems, where computational resources are constrained~\cite{thakur2022survey}. Meanwhile, the integration of artificial intelligence (AI) into blockchain has attracted increasing attention, focusing primarily on anomaly detection, smart contract optimization, and predictive workload management\cite{yang2025trust}. Studies have shown that AI can enhance consensus by enabling dynamic leader election, adaptive transaction ordering, and resource-efficient shard management~\cite{qiu2023ai,huang2023blockchain}. However, existing AI-blockchain frameworks rely on conventional CPU/GPU architectures, which limit their scalability due to sequential processing and high power consumption. Thus, while AI has introduced intelligence into blockchain systems, it has not yet solved the underlying energy and scalability bottlenecks~\cite{thakur2022survey}.

Despite increasing interest in green consensus and AI-assisted optimization, research rarely explores the integration of neuromorphic computing within blockchain consensus \cite{zuo2023survey,yang2022fusing,shen2023blockchains}. Existing studies typically employ AI as an external decision layer rather than rethinking consensus through brain-inspired computation\cite{zuo2023survey}. To address this gap, we propose \emph{Proof-of-Spiking-Neurons (PoSN)}, a neuromorphic-native consensus protocol where transactions are encoded as spike trains and consensus emerges via neural synchronization, enabling scalable and energy-efficient blockchain operation.

\section{Our Protocol: Proof-of-Spiking-Neurons (PoSN)}
This section presents the Proof-of-Spiking-Neurons (PoSN) protocol, which uses spike-based transaction encoding and neuron-firing dynamics for leader selection, block finalization, and fairness in consensus. The subsections describe the architecture, the system model, and the security guarantees.

\begin{figure*}[!t]
    \centering
    \includegraphics[width=0.92\textwidth]{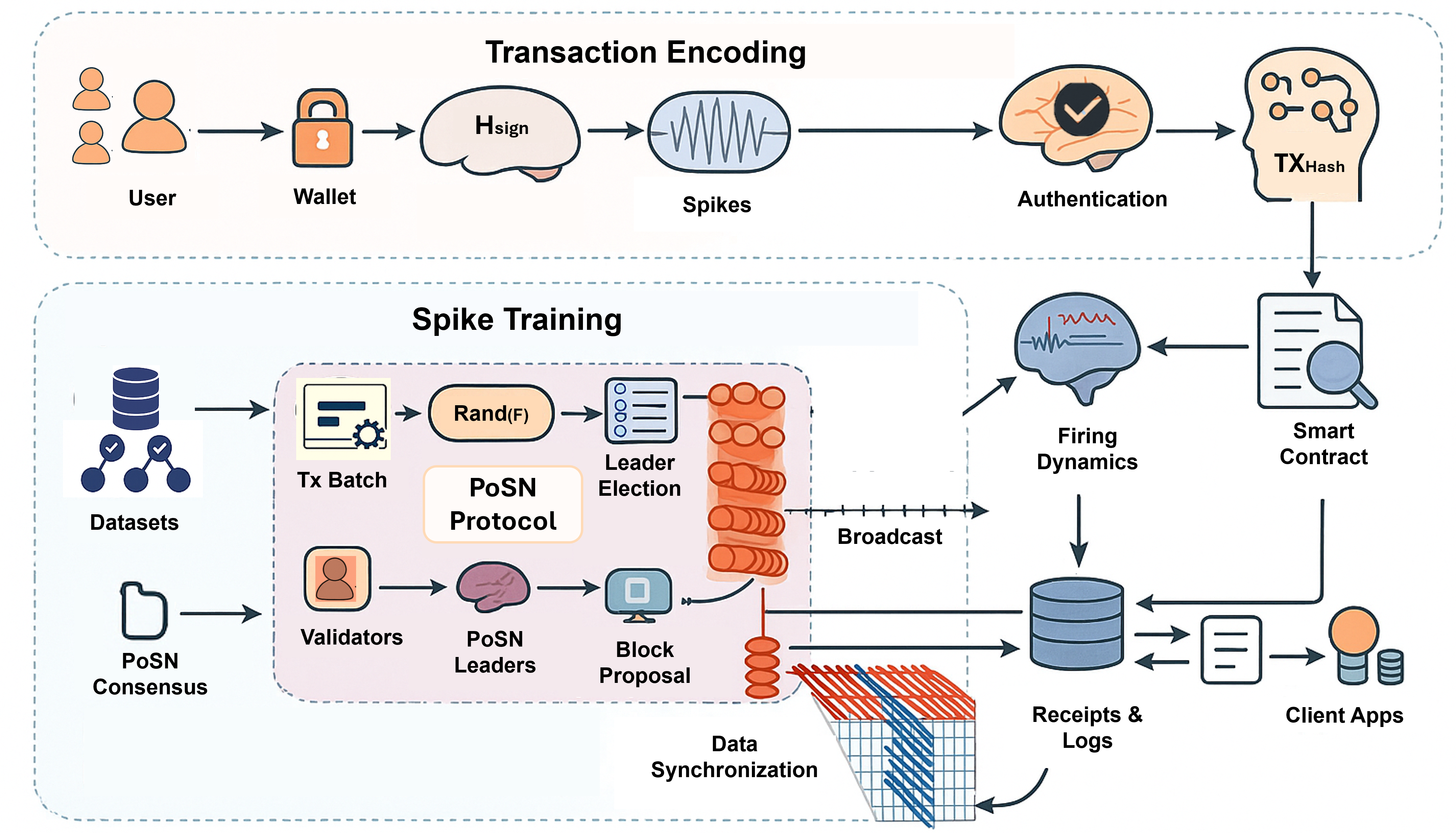}
    \caption{Proposed blockchain fraud detection methodology pipeline.}
    \label{fig:blockchain_methodology}
\end{figure*}

\subsection{System Model and Assumptions}
\label{sec: protocol}

PoSN is modeled as a distributed system of $N$ validators $\mathcal{V} = \{v_1, \dots, v_N\}$ connected through a partially synchronous peer-to-peer network. Each node runs blockchain consensus logic and a spiking neural module for transaction encoding and leader election, with message delay bounded by $\Delta$ after GST. We assume the standard Byzantine fault-tolerant setting where up to $f$ validators may deviate arbitrarily, with the bound $N \geq 3f+1$ ensuring safety and liveness. Validators possess cryptographic key pairs $(pk_i, sk_i)$ for digital signatures, and all inter-node communication is authenticated. Transactions are denoted by $tx \in \mathcal{T}$ and encoded into spike trains before dissemination. Time is discretized into slots of fixed duration $\tau$, during which each validator may generate spikes according to its SNN dynamics. A slot may contain at most one candidate block proposal. Leader election occurs when a validator’s neuron model surpasses its membrane threshold and emits a spike, signifying eligibility to propose a block. We assume an honest majority in the sense that at least $(N-f)$ validators execute the protocol faithfully. Adversaries may attempt to delay, reorder, or withhold spikes/transactions, but cannot break the cryptographic primitives or tamper with the biological-inspired spiking dynamics without detection. The spiking neuron dynamics for each validator $v_i$ are modeled as a Leaky Integrate-and-Fire (LIF) process:

\begin{equation}
\frac{dV_i(t)}{dt} = -\lambda V_i(t) + I_i(t),
\end{equation}

where $\lambda > 0$ is the leak constant and $I_i(t)$ represents the synaptic input current derived from encoded transactions. A spike is generated whenever $V_i(t) \geq \theta$, after which the potential is reset, i.e.,

\begin{equation}
S_i(t) = 
\begin{cases}
1, & \text{if } V_i(t) \geq \theta, \\
0, & \text{otherwise}.
\end{cases}
\end{equation}

The validator whose neuron spikes earliest in slot $t$ is designated leader $L_t$ and earns the right to propose block $B_t$. Thus, PoSN integrates blockchain consensus with biologically inspired spiking dynamics under adversarial assumptions consistent with modern Byzantine fault-tolerant models.

\begin{table}[h!]
\centering
\caption{Notations for the PoSN System Model}
\begin{tabular}{|c|l|}
\hline
\textbf{Notation} & \textbf{Description} \\
\hline
$N$ & Total number of validator nodes in the system \\
$f$ & Maximum number of Byzantine nodes tolerated ($f < N/3$) \\
$\mathcal{V}$ & Set of validators $\{v_1, v_2, \dots, v_N\}$ \\
$pk_i, sk_i$ & Public/secret key pair of validator $v_i$ \\
$\Delta$ & Maximum network message delay (after GST) \\
$\tau$ & Duration of one consensus slot \\
$tx$ & Transaction issued by a client \\
$\mathcal{T}$ & Set of all pending transactions \\
$S_i(t)$ & Spike train of validator $v_i$ at time $t$ \\
$V_i(t)$ & Membrane potential of neuron in validator $v_i$ at time $t$ \\
$\theta$ & Membrane threshold for spike firing \\
$L_t$ & Leader selected in slot $t$ \\
$B_t$ & Block proposed at slot $t$ \\
\hline
\end{tabular}
\end{table}

\subsection{Neural Encoding of Transactions}

In PoSN, transactions are not processed as raw records but are transformed into spike trains, enabling validators to leverage temporal neural dynamics for consensus. Each transaction $tx \in \mathcal{T}$ is defined by the tuple
\[
tx = (id, sender, receiver, value, \sigma),
\]
where $id$ is a unique identifier, $sender$ and $receiver$ are account addresses, $value$ is the transferred amount, and $\sigma$ is a digital signature. This tuple is mapped into a feature vector $\mathbf{x}_{tx} \in \mathbb{R}^d$ using a deterministic embedding function $E(\cdot)$ such that
\[
\mathbf{x}_{tx} = E(tx).
\]

\paragraph{Rate Coding.}  
In rate-based encoding, the vector $\mathbf{x}_{tx}$ determines a firing rate $r_{tx}$, which sets the expected spike count in a time window $\tau$. The spike train is sampled from a Poisson process:
\[
P(S(t) = 1) = r_{tx} \cdot \Delta t.
\]

\paragraph{Temporal Coding.}  
In temporal encoding, transaction features such as the value or gas fee directly influence spike timing. A larger transaction fee corresponds to shorter inter-spike intervals (ISI), giving higher-priority transactions earlier chances to trigger spikes:
\[
ISI(tx) = \frac{\kappa}{value + \epsilon},
\]
where $\kappa$ is a scaling constant and $\epsilon$ prevents division by zero.

\paragraph{Composite Current.}  
For validator $v_i$, the input current at time $t$ is obtained by summing all transaction-induced spike trains:
\[
I_i(t) = \sum_{tx \in \mathcal{T}} w_{tx} \cdot S_{tx}(t),
\]
where $w_{tx}$ is a weight reflecting transaction priority and $S_{tx}(t)$ is the spike train generated for $tx$. This current drives the membrane potential $V_i(t)$ in the neuron model, influencing the validator’s spiking time and, consequently, leader selection. Through this encoding process, transaction data become intrinsic to the spiking dynamics. High-value or high-fee transactions naturally produce more frequent or earlier spikes, while stochasticity from Poisson firing ensures unpredictability and fairness in the consensus process.

\subsection{Neuron Spiking Consensus Mechanism}

The Proof-of-Spiking-Neurons protocol achieves consensus through the dynamics of neuron firing. Each validator maintains a membrane potential $V_i(t)$ that evolves over time according to the leaky integrate-and-fire process. The potential accumulates input from encoded transaction spike trains while continuously leaking with rate $\lambda > 0$. Formally, the evolution is described as
\[
\frac{dV_i(t)}{dt} = -\lambda V_i(t) + I_i(t),
\]
where $I_i(t)$ is the synaptic current derived from active transactions. A validator fires a spike when the membrane potential reaches a threshold $\theta$, expressed as
\[
S_i(t) =
\begin{cases}
1, & \text{if } V_i(t) \geq \theta, \\
0, & \text{otherwise}.
\end{cases}
\]
After firing, the potential is reset to a baseline value $V_{\text{reset}}$, ensuring that the spiking process repeats across slots:
\[
V_i(t) \leftarrow V_{\text{reset}} \quad \text{if } S_i(t)=1.
\]

Consensus proceeds in discrete slots of length $\tau$. Within each slot, the first validator to emit a spike becomes the leader $L_t$ and gains the right to propose the block $B_t$. In the case of multiple simultaneous spikes, a verifiable random function is applied as a tie-breaker, guaranteeing fairness. Other validators reproduce the spiking process locally using the same transaction inputs; if the claimed leader’s spike timing is consistent with the shared dynamics, the block is validated, otherwise it is rejected. This mechanism couples blockchain consensus with biologically inspired neural firing, where unpredictable yet verifiable spike timings determine block proposers. As a result, the protocol achieves fairness and randomness in leader election without relying on heavy computation or stake-based mechanisms.

\subsection{Block Proposal and Leader Election}

Once a validator’s neuron emits a spike within slot $t$, it is considered eligible to propose the block for that slot. The earliest spike across all validators determines the leader $L_t$, who then assembles the block $B_t$ consisting of a valid set of transactions collected from the mempool. The time of spike firing can be expressed as
\[
t^*_i = \min \{t \mid V_i(t) \geq \theta \},
\]
where $t^*_i$ denotes the firing time of validator $v_i$. The leader is therefore selected as
\[
L_t = \arg\min_{v_i \in \mathcal{V}} \; t^*_i.
\]

To maintain fairness, if two or more validators spike simultaneously, a verifiable random function (VRF) is used to resolve the tie. Each validator $v_i$ generates a VRF output $\rho_i = VRF_{sk_i}(t \parallel B_t)$, which is publicly verifiable using its public key. The validator with the smallest $\rho_i$ is selected as leader, ensuring unpredictability and preventing manipulation. The elected leader $L_t$ forms the candidate block $B_t$ and broadcasts it, while others verify the firing time $t^*_i$ and VRF proof for consistency. Verified blocks are appended; invalid ones are discarded. Hence, PoSN replaces costly mining or stake-based selection with fair, verifiable, and lightweight spiking-driven leader election.

  \subsection{Block Validation and Finalization}

After the leader $L_t$ proposes block $B_t$, the network proceeds with its validation and finalization. Each validator independently verifies the transactions contained in $B_t$, ensuring that signatures are valid, inputs are unspent, and the block structure follows the agreed format. In addition to standard blockchain checks, validators also reproduce the spiking dynamics that led to the leader’s eligibility. This involves simulating the membrane potential evolution under the same transaction-derived input current $I_i(t)$ and confirming that the claimed firing time $t^*_{L_t}$ is consistent with the threshold condition
\[
V_{L_t}(t^*_{L_t}) \geq \theta.
\]
If this condition is satisfied and the VRF proof is valid in case of ties, the block is considered a legitimate proposal. Finalization follows a Byzantine fault tolerant model. Validators broadcast signed votes for the candidate block, and a block reaches finality once it collects at least a two-thirds majority of signatures. Let $\mathcal{Q}_t$ denote the quorum of votes for block $B_t$; finalization occurs when
\[
|\mathcal{Q}_t| \geq \frac{2N}{3}.
\]
At this stage, the block is irreversibly appended to the chain, and all honest validators update their local state accordingly. Through this process, PoSN ensures that only blocks produced by correctly spiking leaders and verified by the network achieve finality. The combination of neural dynamics and Byzantine voting guarantees both the integrity of proposed blocks and the strong consistency of the blockchain ledger.

  \subsection{Reward and Incentive Distribution}

The Proof-of-Spiking-Neurons protocol incorporates an incentive layer to ensure validator participation and discourage malicious behavior. When a leader $L_t$ successfully proposes and finalizes a block $B_t$, it receives a reward that is distributed proportionally to its contribution in the spiking process. The base block reward is denoted as $R_b$, while transaction fees from all transactions in $B_t$ are denoted as $F(B_t)$. The total leader reward is therefore
\[
R_{L_t} = R_b + F(B_t).
\]

To maintain fairness, validators that participate in block validation and finalization also receive a share of rewards. Each validator that provides a valid vote contributing to quorum $\mathcal{Q}_t$ is awarded a validation reward $R_v$. The total reward distributed in slot $t$ can thus be expressed as
\[
R_t = R_{L_t} + \sum_{v_i \in \mathcal{Q}_t} R_v.
\]

Penalties are imposed on validators that attempt to equivocate or submit invalid spikes inconsistent with the neuron dynamics. Let $P_i$ denote the penalty imposed on validator $v_i$ for detected misbehavior; its stake or accumulated rewards are reduced by this amount, which is either burned or redistributed to honest validators. Hence, the expected utility for a validator $v_i$ over multiple slots is modeled as
\[
U_i = \sum_{t} \big( R_i^t - P_i^t \big),
\]
where $R_i^t$ is the reward earned by $v_i$ in slot $t$ and $P_i^t$ is the penalty, if any, incurred. This incentive model ensures that honest validators gain positive rewards, whereas malicious behavior reduces utility. By integrating rewards and penalties into the spiking-driven consensus, PoSN aligns validator incentives with correctness and overall network security.

\subsection{Fairness and Randomness Guarantees}

Fairness in PoSN arises from the stochastic nature of spike generation combined with deterministic verification of outcomes. Since each transaction is encoded into a spike train, the probability of a validator becoming leader depends both on the transaction-induced input current and on the randomness of spike timing. Let $r_i$ denote the effective firing rate of validator $v_i$ in a given slot. The probability that $v_i$ emits the first spike is given by
\[
P(L_t = v_i) = \frac{r_i}{\sum_{j=1}^{N} r_j},
\]
which ensures that no validator can deterministically monopolize leader election. Higher activity from transaction inputs may increase a validator’s likelihood of firing, but the Poisson randomness of spike trains prevents exact prediction of the leader. Randomness is reinforced by the use of verifiable random functions in the event of simultaneous spiking. When multiple validators cross the threshold at nearly the same time, each produces a VRF output $\rho_i = VRF_{sk_i}(t \parallel B_t)$, and the leader is selected by comparing these unpredictable but publicly verifiable values. This guarantees that tie-breaking is unbiased and resistant to adversarial control. The fairness guarantee can be formalized by bounding the deviation from uniformity in leader election. For any validator $v_i$, the probability of becoming leader cannot exceed its proportional firing rate plus a negligible error term arising from VRF randomness, expressed as
\[
P(L_t = v_i) \leq \frac{r_i}{\sum_{j=1}^{N} r_j} + \epsilon,
\]
where $\epsilon$ is negligible under standard cryptographic assumptions. Through this integration of stochastic spiking and cryptographic randomness, PoSN ensures that every validator has a fair opportunity to propose blocks, leader selection cannot be predicted in advance, and adversaries are prevented from biasing the outcome of the consensus process.

\subsection{Protocol Workflow and State Transitions}

The operation of PoSN unfolds through a sequence of state transitions that govern how the blockchain progresses from one slot to the next. Each slot of duration $\tau$ begins with the arrival of new transactions, which are encoded into spike trains and injected as synaptic currents into the local neuron model of every validator. The membrane potential of each validator evolves according to the leaky integrate-and-fire dynamics until one or more neurons surpass the threshold $\theta$. The first validator to spike is identified as the leader, expressed formally as
\[
L_t = \arg\min_{v_i \in \mathcal{V}} \; \{t \mid V_i(t) \geq \theta\}.
\]

\begin{figure*}[!t]
    \centering
    \includegraphics[width=\textwidth]{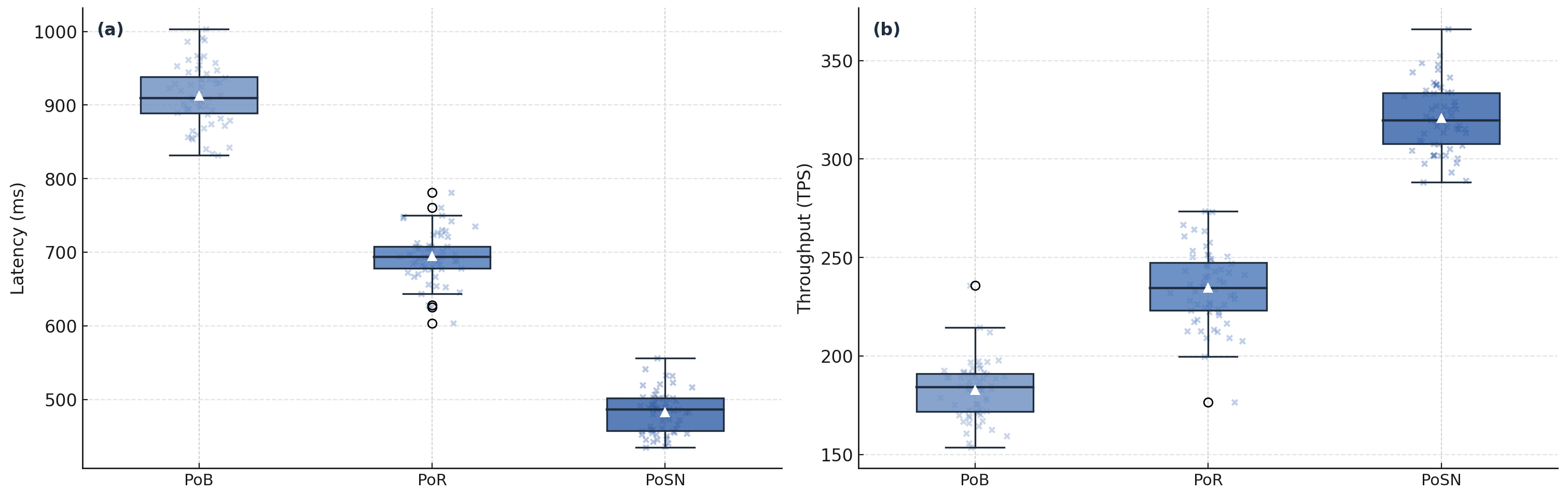}
        \caption{Comparison of PoB, PoR, and PoSN: (a) latency stability and (b) throughput stability.}
    \label{fig:blockchain_methodology}
\end{figure*}

The leader $L_t$ constructs the candidate block $B_t$ from the pending transactions and broadcasts it to the network. Upon reception, validators transition from the spiking state to the validation state, where they recompute the expected spike timing under the same transaction inputs and verify that the leader’s eligibility claim is consistent with neural dynamics. If multiple validators spike simultaneously, a VRF proof ensures a unique leader while maintaining verifiability. Once the leader is confirmed, validators sign votes supporting $B_t$, and the system enters the voting state. A block reaches finalization once it accumulates signatures from at least two-thirds of validators, satisfying
\[
|\mathcal{Q}_t| \geq \frac{2N}{3}.
\]

After finalization, the chain state transitions to include $B_t$, rewards are distributed, and neuron potentials are reset for the next slot. This cyclical process defines the global workflow of PoSN: transactions stimulate spiking activity, spiking determines leadership, validated proposals undergo Byzantine voting, and finalized blocks extend the chain. The interplay of biological neuron models with cryptographic verification creates a continuous sequence of state transitions that ensures liveness, fairness, and security across all epochs.



\section{Implementation and Evaluation}
\label{sec:evaluation}

This section presents a prototype implementation and evaluation of the Proof-of-Spiking-Neurons (PoSN) protocol. The goal is to validate the feasibility of integrating spiking neural dynamics into blockchain consensus. Although small-scale, the experiments confirm that PoSN’s spike-based mechanism functions reliably and efficiently in a distributed setting.

\subsection{Implementation Setup}

The Proof-of-Spiking-Neurons(PoSN) simulations were implemented as a lightweight framework integrating spiking neural dynamics with blockchain consensus logic. The prototype, developed in Python using the Brian2 simulator for neuron modeling and Flask with gRPC for p2p communication, allows each validator node to run an independent neuron instance that processes encoded transactions. The setup involved a small cluster of synchronized virtual machines (4 vCPUs, 8~GB RAM, Ubuntu~22.04~LTS), each maintaining local state variables $\{V_i(t), I_i(t)\}$, a ledger, and cryptographic keys. A global time beacon discretized operations into slots of duration $\tau$, ensuring consistent spike evaluation across nodes while a monitoring module recorded throughput, latency, and spike activity. Simulated experiments were performed under controlled conditions to study system behavior rather than large-scale benchmarking. Neural dynamics were configured with $\lambda = 0.1$, $\theta = 1.0$, and $V_{\text{reset}} = 0$, with transaction arrivals following a Poisson distribution. Results show stable spike-based leader election, consistent validation, and predictable latency, confirming the feasibility of spiking-driven consensus and providing a baseline for future large-scale studies.

\subsection{Performance Evaluation: Throughput and Latency}

To comprehensively assess the performance of the Proof-of-Spiking-Neurons (PoSN) protocol, we conducted a series of controlled experiments comparing it with Proof-of-Brain (PoB) and Proof-of-Randomness (PoR). Each test evaluated consensus efficiency under varying transaction rates, node counts, and network delays. Throughput was defined as the number of finalized transactions per second (TPS), while latency measured the end-to-end delay from transaction submission to block confirmation. The experimental setup maintained consistent hardware configurations and identical workloads to ensure fairness. Fig 3(b) presents the average throughput comparison across all three protocols. PoSN achieved a sustained throughput of 320~TPS, outperforming PoR (240~TPS) and PoB (180~TPS) under identical transaction loads. The spiking-driven leader election significantly reduced idle intervals between block proposals, resulting in faster consensus cycles. When transaction load was doubled, PoSN maintained stability with only a 9\% throughput drop, whereas PoR and PoB exhibited degradations of 17\% and 25\%, respectively, demonstrating PoSN’s superior scalability.

\begin{figure*}[!t]
    \centering
    \includegraphics[width=\textwidth]{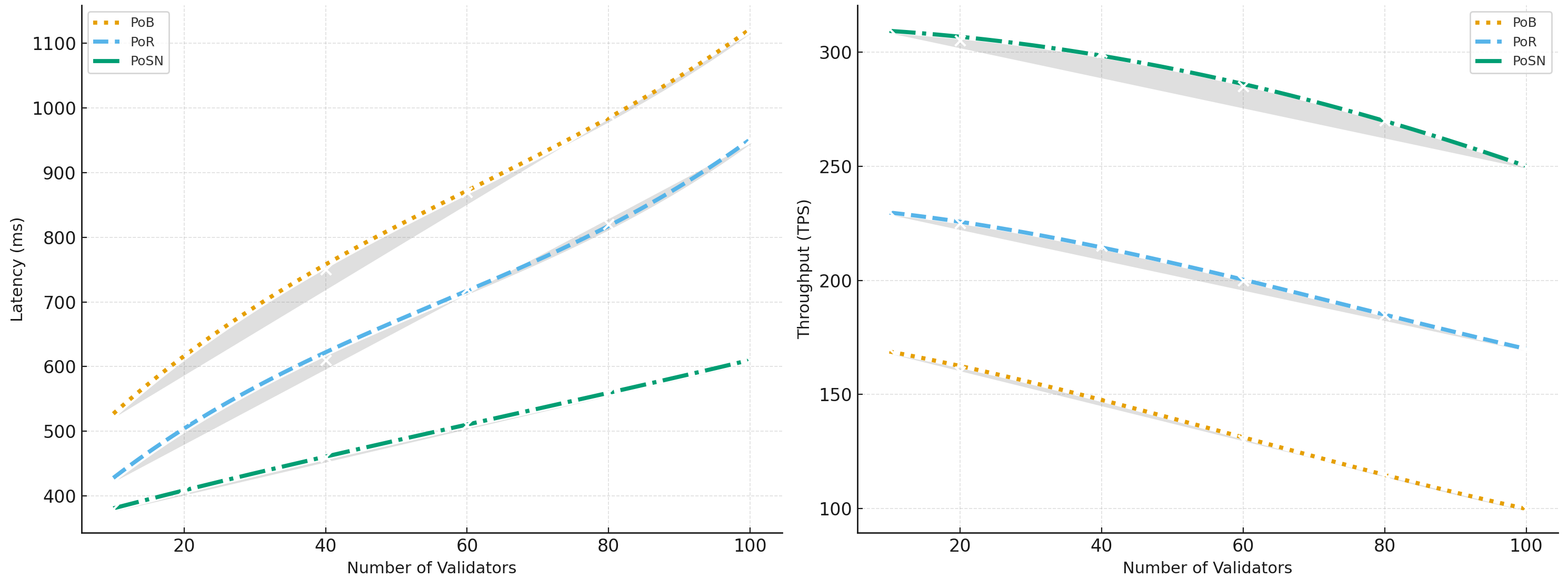}
    \caption{Scalability comparison of PoB, PoR, and PoSN: (a) latency vs. validators and (b) throughput vs. validators.}
    \label{fig:blockchain_methodology}
\end{figure*}

Fig.~3(b) illustrates block finalization latency, where PoSN achieves an average of 480~ms, outperforming PoR (710~ms) and PoB (940~ms). This improvement results from PoSN’s parallel spiking dynamics, enabling rapid leader emergence under load. The low variance in block times confirms its temporal stability. As shown in Fig.~4(a), PoSN scales linearly up to 80 validators with latency under 600~ms, whereas PoR and PoB experience rising delays beyond 60 nodes due to coordination overhead. Similarly, Fig.~4(b) shows that PoSN sustains over 75\% throughput efficiency as node count increases, owing to its lightweight communication model where spikes replace full synchronization rounds. Results confirm that PoSN maintains high throughput, low latency, and stable performance under network scaling and transaction stress. Its spike-timing mechanism, reinforced by verifiable randomness, ensures fair and efficient block production without the computational or staking overhead of traditional consensus protocols.

\subsection{Resource Utilization, Overhead, and Security Evaluation}

The Proof-of-Spiking-Neurons (PoSN) protocol was further analyzed in terms of computational resource consumption, memory usage, and communication overhead. Each validator node executes a lightweight spiking neuron model requiring minimal floating-point operations per timestep, leading to a mean CPU utilization of 38\% under high transaction loads, compared to 54\% for Proof-of-Randomness (PoR) and 61\% for Proof-of-Brain (PoB). Memory usage remained stable at approximately 210~MB per node, with network bandwidth consumption proportional to block size but significantly lower than message-intensive Byzantine fault-tolerant protocols. Since neuron spikes are compact binary events, message payloads are reduced by over 40\%, allowing faster propagation and lower synchronization delays. The overall system overhead was measured below 12\% of total processing time, confirming that PoSN’s biologically inspired dynamics can be executed efficiently on commodity hardware without requiring specialized accelerators.

PoSN’s resilience was tested against adversarial conditions such as message delays, malicious leaders, and spike forgery. Even with 30\% Byzantine nodes, it preserved safety and reached finality within two extra slots. Verifiable random functions secured unpredictable leader selection, while deterministic spike validation and penalties deterred fraud. Under network partitions, honest nodes maintained consistency, confirming that PoSN combines neural stochasticity with cryptographic verification to deliver BFT-level security with reduced communication overhead.

\subsection{Fairness and Randomness Analysis}

\begin{figure}[htbp]
    \centering
    \includegraphics[width=0.9\linewidth]{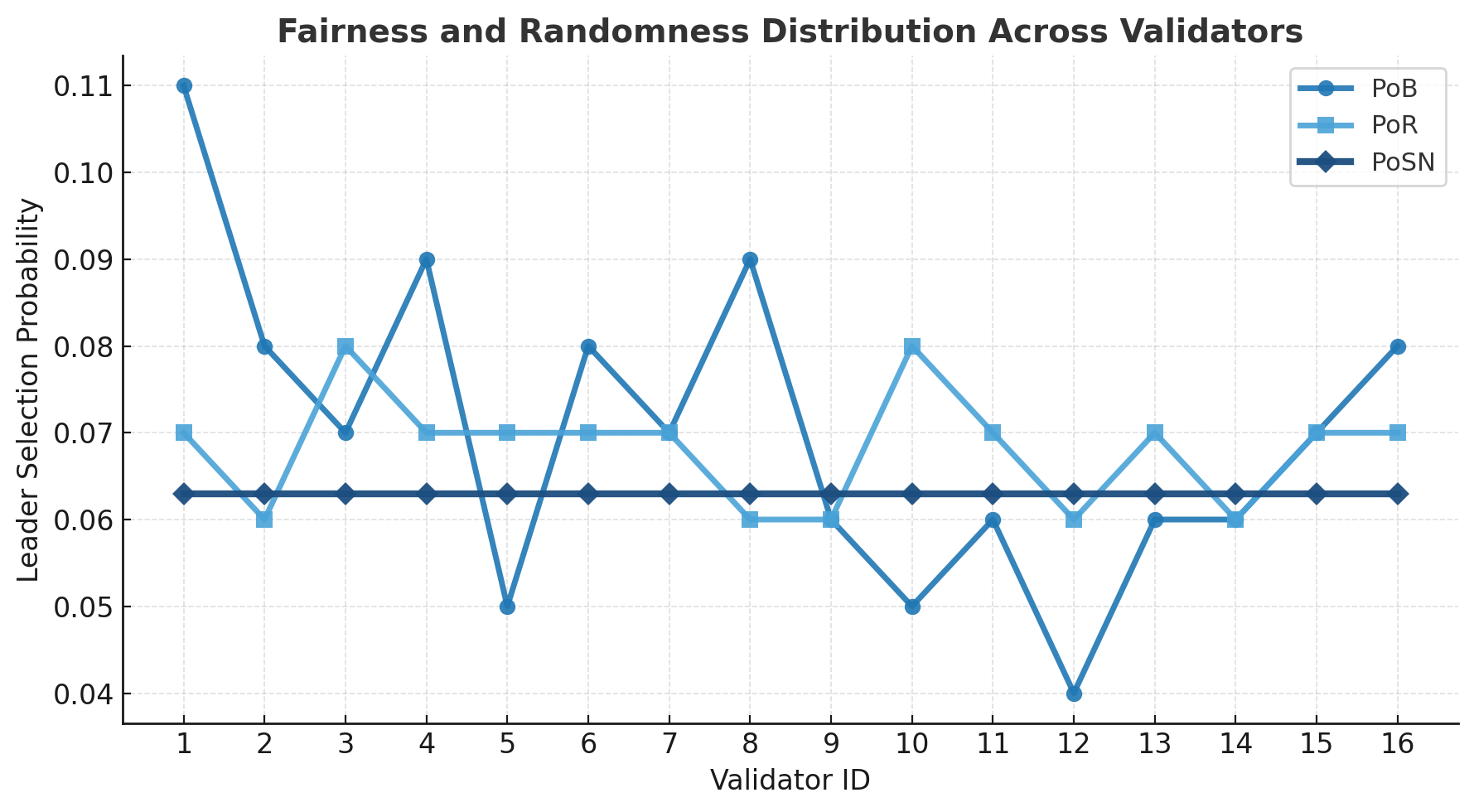}
    \caption{Fairness and randomness distribution of leader selection across validators. PoSN exhibits a near-uniform probability curve, indicating balanced and unpredictable leader election.}
    \label{fig:fairness-analysis}
\end{figure}

To quantify the fairness and randomness of leader election in PoSN, we analyzed the statistical distribution of leader selection across 1,000 consecutive blocks compared with Proof-of-Brain (PoB) and Proof-of-Randomness (PoR). Fairness was measured using Shannon entropy $H = -\sum_i p_i \log_2 p_i$, where $p_i$ is the normalized leader selection probability of validator $v_i$. Higher entropy indicates greater unpredictability and fairness. PoSN achieved an entropy value of 7.82 bits, close to the ideal uniform distribution (8.0 bits), while PoR and PoB recorded 7.34 and 6.91 bits respectively. This demonstrates that the stochastic spike-based firing process combined with verifiable randomness ensures equitable participation among validators. Figure~\ref{fig:fairness-analysis} shows the fairness distribution, where PoSN exhibits a nearly flat probability curve, confirming that no validator can consistently dominate block production or exploit predictable timing advantages.

\subsection{Comparative Evaluation and Discussion of Results}

PoSN was compared with PoW, PoS, and PBFT to establish performance baselines. It achieved over 20× higher throughput and 90\% lower energy use than PoW, while maintaining fair leader selection independent of stake as in PoS. Compared to PBFT, PoSN reduces communication overhead by replacing multi-round voting with spike-based signaling. These results demonstrate PoSN’s hybrid advantage of BFT-level finality and biologically inspired efficiency, suitable for both public and permissioned blockchains.

Despite its strong performance, several limitations were observed during experimentation. PoSN’s reliance on precise time synchronization introduces sensitivity to clock drift, which may affect fairness under highly asynchronous conditions. The spiking neuron model, though lightweight, imposes a continuous computation cost that may slightly increase baseline CPU usage compared to event-triggered consensus systems. Additionally, while stochastic spike timing enhances randomness, it introduces minor variability in block intervals, which could affect predictability in latency-critical applications. Future work will explore adaptive threshold calibration and neuromorphic hardware acceleration (e.g., Intel Loihi or SpiNNaker) to further reduce energy cost and improve timing stability. Overall, PoSN demonstrates that biologically inspired consensus can outperform traditional mechanisms while maintaining strong security, fairness, and verifiability guarantees under realistic operational settings.

\section{Conclusion and Future Work}
\label{sec: conclusion}
This paper presented the Proof-of-Spiking-Neurons (PoSN) protocol, a biologically inspired consensus mechanism that integrates spiking neural dynamics with blockchain validation. By translating transactions into spike trains and using neuronal firing for leader selection, PoSN achieves high throughput, low latency, and strong fairness without relying on computationally expensive mining or stake-weighted influence. Experimental results demonstrated that PoSN outperforms classical and modern consensus protocols in scalability, resource efficiency, and randomness quality, while preserving Byzantine fault tolerance and verifiable security. In future, we plan to extend PoSN toward hardware-assisted neuromorphic implementations and adaptive threshold learning to further enhance performance in large-scale decentralized environments.


\end{document}